\documentclass[onecolumn]{emulateapj}

\def\LL{{\cal L}}

\shorttitle{Stellar radii from asteroseismic data}
\shortauthors{Basu et al.}

\begin{document}


\title{Determination of stellar radii from asteroseismic Data}

\author{Sarbani Basu}
\affil{Department of Astronomy, Yale University, P.O. Box 208101, New
Haven, CT 06520-8101; sarbani.basu@yale.edu}

\author{William J. Chaplin, Yvonne Elsworth}
 \affil{School of Physics
and Astronomy, University of Birmingham, Edgbaston, Birmingham B15
2TT, U.K.; w.j.chaplin@bham.ac.uk, y.p.elsworth@bham.ac.uk}

\begin{abstract}

The NASA Kepler mission is designed to find planets through
transits. Accurate and precise radii of the detected planets depend on
knowing the radius of the host star accurately, which is difficult
unless the temperature and luminosity of the star are known
precisely. Kepler, however, has an asteroseismology programme that
will provide seismic variables that can characterise stellar radii
easily, accurately, and extremely precisely.  In this paper we
describe the Yale-Birmingham (YB) method to determine stellar radii using a
combination of seismic and conventional variables, and analyse the
effect of these variables on the result.  We find that for
main-sequence stars, a knowledge of the parallax is not important to get
accurate radii using the YB method: we
can get results to an accuracy and precision of better than a few
percent if we know the effective temperature and the seismic
parameters for these stars. Metallicity does not make much difference
either. However, good estimates of the effective temperature and metallicity,
along with those of the seismic parameters, are essential to determine
radii of sub giants properly. On the other hand, for red giants we
find that determining radii properly is not possible without a good
estimate of the parallax. We find that the so called ``surface term''
in the seismic data has minimal effect on the inferred
radii. Uncertainties in the convective mixing length can matter under
some circumstances and can cause a systematic shift in the inferred
radii.  Blind tests with data simulated to match those expected from
the asteroseismic Survey Phase of Kepler show that it will be possible
to infer stellar radii successfully using our method.

\end{abstract}


\section{Introduction}
\label{sec:intro}

Asteroseismology of solar-type stars promises significant improvements
in our understanding of stellar evolution theory.  By using
asteroseismic data on stars showing solar-like oscillations, in
combination with more traditional non-seismic data, it is
possible to constrain the fundamental stellar parameters to levels of
precision that would not otherwise be possible. There are strong
synergies with exoplanet studies, in that the asteroseismic data can
place tight constraints on the ages and angular momenta of
exoplanetary systems, and also help to place tight constraints on the
exoplanet radii (e.g., see Kjeldsen et al. 2009).

Accurate, high-precision results on fundamental parameters of stars
showing solar-like oscillations have already been demonstrated with
asteroseismic data from ground-based campaigns (e.g., Bedding et
al. 2007). Long (multi-month) high-quality datasets from the CoRoT
satellite (Baglin et al. 2006) are now being exploited for a selection
of F-type main-sequence stars showing solar-like oscillations (e.g.,
Michel et al. 2008; Appourchaux et al. 2008). But it is with the NASA
Discovery-class Kepler mission (Borucki et al. 2008) that we will be
able to perform a proper ``seismic survey'' of the solar-like part of
the color-magnitude diagram, thanks to the large number of stars it
will observe for asteroseismology.

Kepler was launched successfully on 2009 March 7.  In addition to
searching for Earth-like exoplanets via the transit method, the Kepler
Asteroseismology Science Consortium
(KASC)\footnote{http://kepler.asteroseismology.org} will have an
unprecedented opportunity to study more than one-thousand stars
showing solar-like oscillations (Christensen-Dalsgaard et al. 2008a,
b). This large volume of data will be collected during the initial
Survey Phase, when the solar-like targets will each be observed for
one month. These targets have been selected from the Kepler Input
Catalogue (KIC; see Brown et al. 2005).  On completion of the Survey
Phase, a subset of approximately 50 to 75 solar-like targets will be
selected for continuous multi-year observations, ideally through the
duration of the rest of the mission.

The asteroFLAG consortium (Chaplin et al. 2008a) has been at the
forefront of developing semi- and fully-automated analysis pipelines
for the solar-like Kepler data. An automated approach is strongly
indicated, due to the large number of Survey Phase targets. Stello et
al. (2009) described a variety of pipeline approaches for Kepler, most
of them automated, to constrain the radii of solar-like stars using
the principal frequency spacings of the oscillation spectra as the
seismic inputs. The pipelines were tested on artificial F, G and K
main-sequence stars, in a series of asteroFLAG hare-and-hounds
exercises.

In this paper, we describe our pipeline, the Yale-Birmingham (YB)
pipeline, for determining stellar radii using a mixture of seismic
inputs and conventional stellar parameters. We test the pipeline using
new stellar models, in addition to updated versions of the asteroFLAG
hare-and-hounds models; and use an additional seismic input not
adopted previously, the frequency of maximum oscillation power.  This
paper also adopts levels of precision expected for the one-month
Survey Phase data. (Detailed information on the Survey Phase strategy
was not to hand when much of the work for Stello et al.  was
conducted.) We also analyse the effect on the inferred radii of the
sizes of the errors in the different inputs.  A systematic study of
this type, which incorporates non-seismic \textit{and} seismic inputs,
has not been presented in detail before and is needed to determine
which non-seismic parameters are most important for an accurate radius
determination.  We test our pipeline on artificial stars at all stages
of evolution --- from the main sequence to the red giant branch.

We have organised the rest of the paper as follows. We discuss how we
determine stellar radii in Section~\ref{sec:method}, and we evaluate
the effect of the errors on input variables in
Section~\ref{sec:test}. In Section~\ref{sec:sys} we analyse some known
sources of systematic errors. We show results from a blind test in
Section~\ref{sec:hh}. Finally, we present our conclusions in
Section~\ref{sec:disc}.

\section{Method}
\label{sec:method}

Analysis of data on each solar-like Kepler target will provide a set
of observational input parameters. We consider the case where two
seismic input parameters are available: the so-called large frequency
spacing, $\Delta\nu$, and the frequency of maximum oscillation power,
$\nu_{\rm max}$. We discuss these seismic parameters below. We also
assume the following non-seismic parameters are available: the
effective temperature, $T_{\rm eff}$, the luminosity (or rather the
apparent magnitude, $V$, and parallax, $\pi$), and the metallicity
$Z/X$.  When the first solar-like analyses are performed on Kepler
data, for most stars these non-seismic data will come from the Kepler
Input Catalogue (KIC) (note that parallaxes, from Kepler astrometry,
will follow later). Unlike some of the methods discussed in Stello et
al.~(2009), we do not use $\log g$ since estimates of this quantity
can be unreliable. However, as can be seen later, we could easily add
that variable to the list.

The frequency power spectra of oscillations in solar-type stars
present a pattern of peaks with near regular spacings.  The signatures
of these spacings are quite amenable to extraction, owing to their
regularity. The large frequency spacings, $\Delta\nu$, are the
spacings between consecutive overtones, $n$, having the same spherical
angular degree, $l$, and are related to the acoustic radii of the
stars. When the signal-to-noise ratios in the seismic data are
insufficient to allow robust extraction of individual oscillation
frequencies, it is still possible to extract estimates of the large
frequency spacings for use as the seismic input data. We design our
method for this eventuality, using an average value of the spacing as
the principal seismic input. This is all that we can expect to get for
many of the main-sequence targets observed by Kepler in its Survey
Phase. The large spacing is formally related to the mean density of a
star (see e.g., Christensen-Dalsgaard 1993). It scales as
$(M/R^3)^{1/2}$ where $M$ is the total mass and $R$ the radius of the
star.

If available we also use $\nu_{\rm max}$, the frequency of maximum
power in the oscillation power spectrum. The frequency of maximum
oscillation power is related to the acoustic cut-off frequency of a
star (e.g., see Kjeldsen \& Bedding 1995; Bedding \& Kjeldsen 2003;
Chaplin et al. 2008b), which in turn scales as $M\,R^{-2}\,T_{\rm
eff}^{-1/2}$, where $T_{\rm eff}$ is the effective temperature, and
hence contains information about a star's radius.

In order to estimate the radius, given the input parameters, we make
use of a large, fixed grid of stellar evolutionary models.  Our method
is based on finding, in some sense, the maximum likelihood of the set
of input parameters calculated with respect to the grid of models.
The likelihood function is formally defined as
\begin{equation}
\LL(T_{\rm eff},\log(Z/X), V,\pi,\Delta\nu,\nu_{\rm max})=\LL_{T_{\rm eff}}\;\LL_{\log(Z/X)}\;\LL_{V,\pi}\;\LL_{\Delta\nu}\;
\LL_{\nu_{\rm max}},
\label{eq:like}
\end{equation}
where
\begin{equation}
\LL_{T_{\rm eff}}={1\over\sqrt{2\pi}\sigma(T)} \exp-\left({(T_{\rm obs}-T_{\rm model})^2\over{2\sigma(T)^2}}\right),
\label{eq:likeT}
\end{equation}
with $T=T_{\rm eff}$,
\begin{equation}
\LL_{\log{Z/X}}={1\over\sqrt{2\pi}\sigma(\log(Z/X))} \exp-\left({(\log(Z/X)_{\rm obs}-\log(Z/X)_{\rm model})^2\over{2\sigma(\log(Z/X))^2}}\right),
\label{eq:likeZ}
\end{equation}
\begin{equation}
\LL_{V,\pi}={1\over\sqrt{2\pi}\sigma(\pi)} \exp-\left( {\left( \pi_{\rm obs} - 10^{\left({M_V-V_{\rm obs}\over 5} -1\right)}\right)^2
\over{2\sigma(\pi)^2}}\right),
\label{eq:likepi}
\end{equation}
where in Eq.~\ref{eq:likepi} above we have assumed at all errors are
due to errors in $\pi$ (adding errors in $V$ can be done easily),
\begin{equation}
\LL_{\Delta\nu}={1\over\sqrt{2\pi}\sigma(\Delta\nu)} \exp-\left( ( \Delta\nu_{\rm obs} - \Delta\nu_{\rm model} )^2
\over{2\sigma(\Delta\nu)^2}\right).
\label{eq:likedel}
\end{equation}
and finally
\begin{equation}
\LL_{\nu_{\rm max}}={1\over\sqrt{2\pi}\sigma(\nu_{\rm max})} \exp-\left( ( \nu_{\rm max, obs} - \nu_{\rm max, model} )^2
\over{2\sigma(\nu_{\rm max})^2}\right),
\label{eq:likemu}
\end{equation}
As can be seen from the form of the likelihood function in
Eq.~\ref{eq:like}, we can easily include more variables, or drop other
variables.

\begin{figure*}
\epsscale{0.9}
\plotone{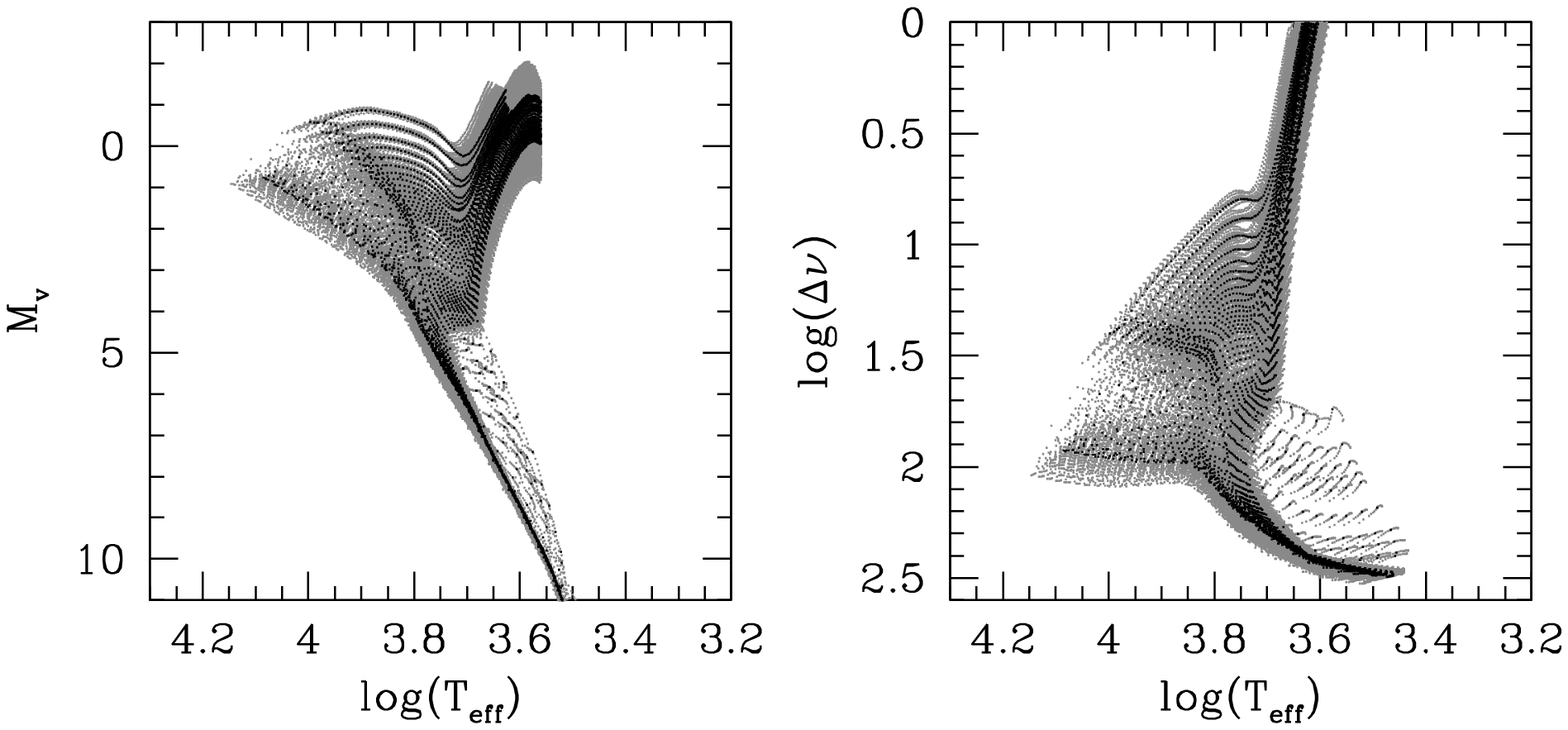}
\caption{The grid of models used in this work to determine stellar
radii. Models for [Fe/H]$=0$ are in black. We show the grid on a
conventional HR diagram (left panel), as well as on the $\log T_{\rm
eff}$-$\Delta\nu$ plane (right panel).}
\label{fig:hrd}
\epsscale{0.5}
\plotone{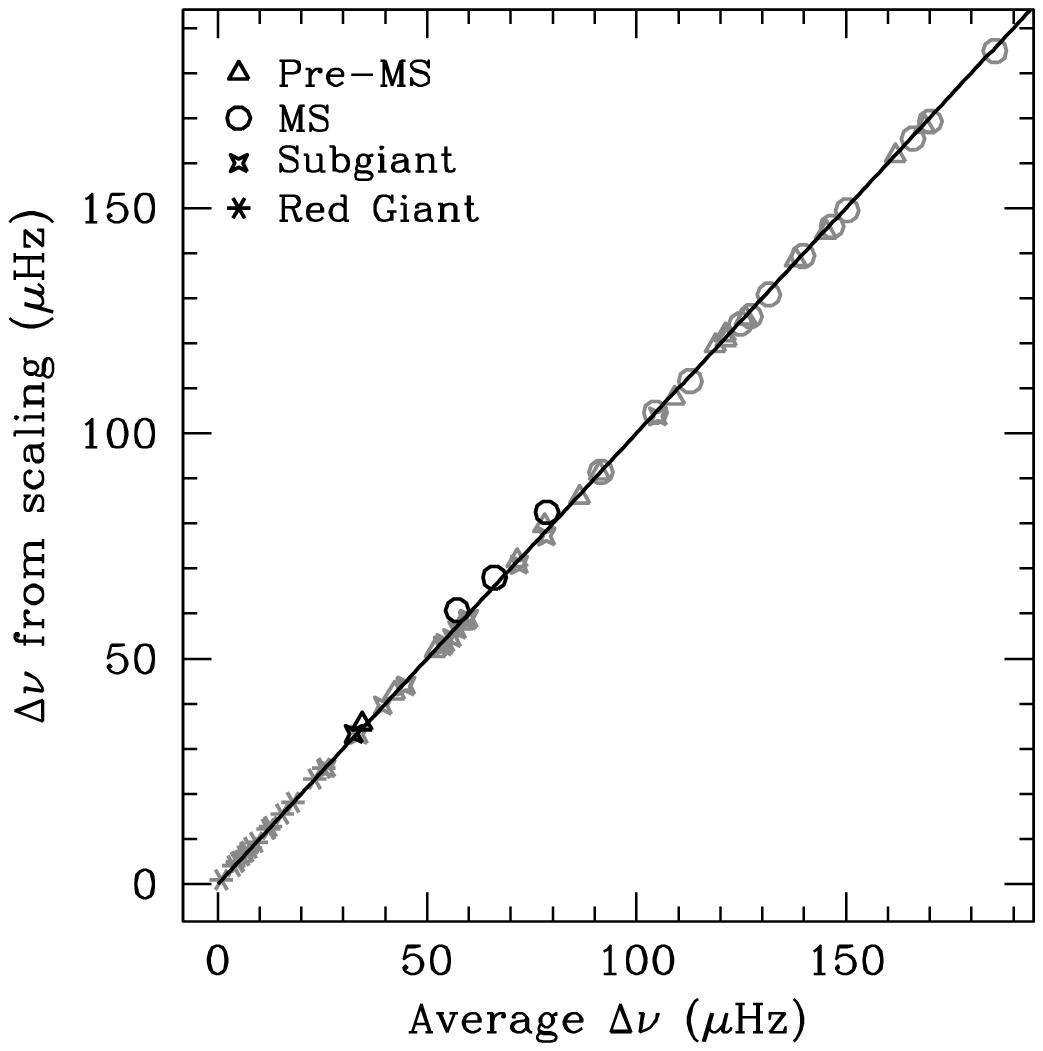}
\caption{Large spacings obtained by scaling from the solar value,
plotted as a function of the average large spacing calculated after
computing the individual oscillation frequencies. Points in grey are
stellar models with $T_{\rm eff}\le 6750$K while those in black have
$T_{\rm eff} > 6750$K. The straight line is
the $x=y$ curve and matches the points extremely well. Only the hotter
models in the sample seem to deviate from the relation.
sample. The evolutionary state does not appear to change the
relationship between the scaled and calculated large separations.}
\label{fig:scale}
\end{figure*}

The grid of evolutionary models used are from the Yale-Yonsei (YY)
isochrones (see Demarque et al.~2004 and references therein).  The
models were constructed with a mixing length parameter of
$\alpha=1.7431$, OPAL high temperature opacities (Rogers \& Iglesias
1995; Iglesias \& Rogers 1996) and Alexander \& Ferguson (1994) low
temperature opacities.  The OPAL equation of state (Rogers et
al. 1996) was used. The models include diffusion. Core overshoot of
$0.2H_p$ is included. The models assume the solar mixture of Grevesse
\& Noels (1993) and the YY solar model has $(Z/X)_\odot=0.0243$ and
$Y=0.2356$. Higher $Z$ models were made assuming $\Delta Y/\Delta
Z=2$. We use the models where the conversion of luminosities to
absolute visual magnitude was made using the colour tables of Lejeune
et al.~(1997).

\begin{figure}
\epsscale{0.4}
\plotone{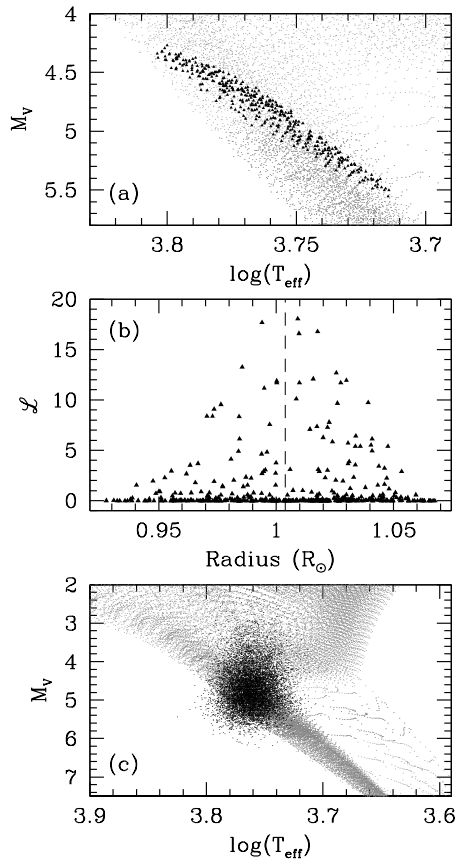}
\caption{(a) Models that contribute to the likelihood function when
the central parameter values for BiSON data are used (black
points). Points in grey show rest of the models in the grid.  (b) The
value of the likelihood function for BiSON data plotted as a function
of radius. The vertical line marks the centre of gravity of the
function. (c) The 10000 models obtained by adding Gaussian errors to
the central BiSON values. The centre of gravity of likelihood function
of each model forms the distribution function we use to determine the
final value of the radius, and errors therein.}
\label{fig:likely}
\end{figure}

The grid of YY models that we have used have [Fe/H] of 0, $\pm 0.05$,
$\pm 0.1$, $\pm 0.15$, $\pm 0.20$, $\pm 0.25$, $\pm 0.3$, $\pm 0.4$,
and $\pm 0.5$ (where ${\rm [Fe/H]} = \log(Z/X) - \log(Z/X)_{\odot}$).
The helium abundance of these models were determined assuming $\Delta
Y/\Delta Z=2$.  We have also restricted the grid to models with ages
between $0.02$ and $14\,\rm Gyr$. The lower age limit ensures that we
do not have too many pre-main sequence models, although we do not
avoid them completely.  We show the grid we use in
Figure~\ref{fig:hrd}. There are about 95000 models in the grid.

The average large frequency spacings $\Delta\nu$ of these models
were estimated by simply scaling from the average solar value, using
the aforementioned $(M/R^3)^{1/2}$ scaling. The alternative would be
to first calculate the pulsational frequencies of each model, and from
these determine an average large frequency spacing. However, it turns
out that the simple scaled values are more than adequate for this
work, as can be seen from Figure~\ref{fig:scale} where we compare for
a representative subset of models the large spacing obtained from
the scaling relationship to the actual average large spacing from
the calculated frequencies. The scaling relation clearly works well. A
small deviation is seen for some of the hotter models. This does not
concern us much since such hot stars are unlikely to show pulsations.

When we also made use of the frequency of maximum oscillation power,
$\nu_{\rm max}$, as part of the input parameter set we estimated its
value for each model by again scaling from the average solar value of
$3100\,\rm \mu Hz$ using the aforementioned $M\,R^{-2}\,T_{\rm
eff}^{-1/2}$ scaling.

The first key step of our method is based on generating 10,000 input
parameter sets by adding different random realisations of Gaussian
noise to the actual (central) observational input parameter set
described at the beginning of this section. This is to allow us to
make a proper determination of the errors involved. The use of these
randomly perturbed parameter sets also allows us to account for the
fact that the relation between the radius of a star and its other
properties, such as $T_{\rm eff}$, metallicity and luminosity, is
extremely non-linear. Thus, while the 10,000 perturbed parameter sets
are created by adding Gaussian noise, the distribution of radii
obtained from the perturbed sets does not usually have a Gaussian
distribution.  The distribution of radii obtained from the central
parameter set and the 10,000 perturbed parameter sets is what we refer
to in the rest of the paper as the ``distribution function.''  We
settled on the number 10,000 as a compromise between obtaining an
eventual smooth distribution of radii in a reasonable computational
time, given one set of input variables.

For the central input parameter set and each of the perturbed sets we
calculate the likelihood, $\LL$, for models within 3$\sigma$ of the
variables of each set. Note that we only choose models that have
characteristics within 3$\sigma$ of {\it all} the variables.  We have
a pre-defined minimum over which we calculate the likelihood to avoid
problems associated with a discrete grid of models.  We find that the
$\pm 3\sigma$ limit is sufficient, with the added benefit that it
saves on computer time, since fewer models are involved (e.g., results
obtained using $\pm 3\sigma$ limits lie well within the errors of
those given for the $\pm 4\sigma$ limits). For the cases we have described below,
a few hundred of the 95000 models contribute to the likelihood function
when seismic constraints are present. In the absence of seismic constraints, the
number of models that contribute to the likelihood
function rises to a few thousand.

The likelihood can be described as a function of the radius of the
model that is used to calculate it. Since we use a discrete grid of
models, we obtain a discrete likelihood distribution function, i.e., a
distribution function made up of calculated values of $\LL$ for each
of the tested models from the discrete model grid. Fitting a
functional form and then determining the maximum likelihood is time
consuming, and using the maximum of the distribution function of $\LL$
gives biases that vary across the HR diagram. We therefore define the
inferred radius of a given set of inputs to be the centre of gravity
of the distribution function.  Once we have inferred the radius for
each of the 10,001 variable sets (again: one central set and 10,000
perturbed sets), we can plot the distribution function of the radii
(see below). We choose the second quartile point of the function as
the final value of the radius and the inter-quartile distances as the
errors on the solution.

To test whether our method works, we first tried to determine the
solar radius using the solar large spacing obtained from solar
$\ell=0$ frequencies measured by the Birmingham Solar-Oscillations
Network (Chaplin et al. 1996). In particular we use the mode set
BiSON-1 described in Basu et al. (2009) to determine an the average
large spacing calculated between 2.47\,mHz and 3.82\,mHz for use as a
seismic input parameter.  The interval was chosen to be roughly ten
large spacings centred around the frequency of maximum power. This
choice is prompted by what we can expect from Kepler in the Survey
Phase.  For the purposes of determining likelihood functions $\LL$,
and subsequently the distribution function of these $\LL$, we assumed
that the Sun has a parallax of $8.9\pm 2.2$\,mas, and that the errors
in the effective temperature, $\log(Z/X)$ and $\Delta\nu$ are 200\,K,
0.2\,dex and 0.56\,$\mu$Hz, respectively. We also conducted a second
exercise were we reduced all the errors by a factor of two.  These
errors are representative of a ``typical'' solar-like main-sequence
target for Kepler (see also Section~\ref{sec:hh} below). Computed
parallaxes of our tests were fixed according to the known apparent
magnitude range for selected solar-like Survey Phase targets, i.e.,
roughly $V=8$ to $V=11.5$.  Note that the precision in the parallax
has been chosen to be similar to that in the existing Hipparcos
catalogue (at the same apparent magnitude), very likely a pessimistic
estimate of the precision we might expect to obtain from Kepler
astrometry.

\begin{figure}
\epsscale{0.5}
\plotone{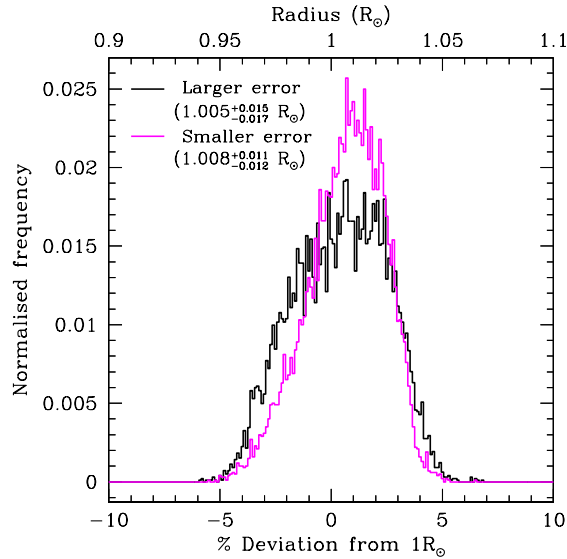}
\caption{Results obtained in our attempt to infer the solar radius
using the large frequency spacings from BiSON data. We plot the
distribution function as a function of the relative error (in \%) of
the results for two different error distributions. The `larger' error
distribution assumes that $\sigma(T_{\rm eff})=200$K,
$\sigma(\log(Z/X)=0.2$, $\sigma(\pi)=2.2$mas,
$\sigma(\Delta\nu)=0.56\mu$Hz and $\sigma(\nu_{max})=125\mu$Hz.  The
`smaller' error-distribution assumes that the errors in each of the
parameters was exactly half that in the `large' distribution.  Note
that the distributions are asymmetric, a result of the highly
non-linear relationship of stellar radii to their other
properties. The inferred solar radius from the two distributions is
noted in the figure. As can be seen, the statistical errors are much
larger than the systematic error in the results.}
\label{fig:bison}
\end{figure}
Since our initial aim was to check whether the method works, rather
than to assess the effect of errors, we did not add random errors to
the input parameters (i.e., the true, underlying input parameters),
which would have more closely mimiced the case of real
observations. The estimated errors on the input parameters were used
only to choose the grid models for the likelihood function, and to
generate the randomly perturbed models (see above) used to calculate
the distribution function.  Section~\ref{sec:hh} describes the result
of an exercise where data with added random errors were used as
inputs.  Figure~\ref{fig:likely}(a) shows the models of our grid that
were used to calculate the likelihood function for the central BiSON
values, in the larger error case; Figure~\ref{fig:likely}(b) shows the
values of the likelihood function against the radius of the models. 390 
models contribute to the likelohood function. As
can be seen, the function is well peaked, however, the discrete nature
of the grid makes the process of finding the location of the maximum
rather unstable (and stable ways are time-consuming and unpractical
given, the large volumes of real Kepler data we expect), hence we
choose to use the centre of gravity of the distribution, which in the
figure is indicated by the vertical line. Figure~\ref{fig:likely}(c)
shows the 10,000 models generated to construct the distribution
function. The distribution function is constructed by first
calculating the likelihood function for each of the models and finding
the centre of gravity of each function.

The two distribution functions are plotted in
Figure~\ref{fig:bison}. We have plotted the distribution functions
against the deviation from $1\,R_\odot$ to give a clearer
understanding of the errors. The radius can be read from the top of
the figure. The first thing to note is that the distribution functions
are definitely not Gaussian, and are asymmetric. This is expected
given the non-linear relationship between the radius of a star and its
temperature, metallicity and luminosity. The solar radius we obtain
from this exercise is $1.005^{+0.015}_{-0.017}R_\odot$ for the large
error case and $1.008^{+0.011}_{-0.012}$ for the smaller error case.
As mentioned earlier, the solution is the second quartile point of the
distribution and the errors represent the distance to the first and
third quartile points.  Thus we are able to determine successfully the
solar radius, demonstrating that the method works. In the following
section we examine the behaviour of the distribution function for
stars at various stages of their evolution and the effect of the input
errors on the results.

\section{Systematic tests of the algorithm}
\label{sec:test}

To determine what parameters affect radius determination the most, we
performed a series of tests with different models. The models are
listed in Table~\ref{tab:teststars}. We consider models at six
different positions on the HR diagram, starting at the lower main
sequence, right up to the red giant branch. The models used in this
section are similar to the calibration models used. The effect of
having different models is shown in subsequent sections. We use this
section to examine the r\^ole of the various input parameters and the
errors on those parameters.

\begin{deluxetable*}{lccccrrl}
\tablecolumns{8}
\tablecaption{Characteristics of the models used to test the radius-finding technique}
\tablehead{\colhead{Name}& \colhead{$Z/X$} & \colhead{$T_{\rm eff}$} &  \colhead{$M_{\rm V}$}
 &  \colhead{$\Delta\nu$} &  \colhead{$\nu_{\rm max}$} & \colhead{Radius} &  \colhead{Comments}\\
\colhead{\ }& \colhead{\ }& \colhead{(K)} & \colhead{\ } & \colhead{($\mu$Hz)} & 
\colhead{($\mu$Hz)} & \colhead{(R$_\odot$)}& }
\startdata
Star 1 &  0.020 &  4530 & 7.38 & 225.81             & 6164              & 0.62 & Main sequence star \\
Star 2 &  0.051 &  5175 & 5.84 & 172.41             & 4517              & 0.83 & Main sequence star \\
Star 3 &  0.036 &  5778 & 4.85 & 142.15             & 3456              & 0.99 & Near turn-off \\
Star 4 &  0.025 &  6372 & 3.17 & {\phantom{1}}70.52 & 1396              & 1.71 & Almost sub giant \\
Star 5 &  0.013 &  6159 & 2.27 & {\phantom{1}}33.73 & {\phantom{1}}541  & 2.84 & Sub giant \\
Star 6 &  0.051 &  4410 & 2.20 & {\phantom{10}}2.20 & {\phantom{1}}21  & 21.44 & Red giant \\
\enddata
\label{tab:teststars}
\end{deluxetable*}
\begin{deluxetable}{cccc}
\tablecaption{Assumed errors on stellar parameters}
\tablehead{\colhead{\ }
    & \colhead{Error 1} & \colhead{Error 2} & \colhead{Error 3}}
\startdata
Assumed parallax (mas) & 18.5 & 8.9 & 5.4 \\
\hline
Parameter & & 1$\sigma$ error in parameter & \\
\hline
$T_{\rm eff}$ (K) & 200 & 200 & 200 \\
$\log(Z/X)$       & 0.2 & 0.2 & 0.2 \\
$\Delta\nu$ ($\mu$Hz) & 0.30 & 0.56 & 1.89 \\
$\nu_{\rm max}$ ($\mu$Hz) & 80 & 125 & 285 \\
$\pi$ (mas) & 1.2 & 2.2 & 3.6 \\
\enddata
\label{tab:error}
\end{deluxetable}

We use three sets of errors in the input parameters, which are listed
in Table~\ref{tab:error}.  The errors in the non-seismic parameters,
particularly $T_{\rm eff}$ and $Z/X$, are motivated by what are
believed to be typical errors in the Kepler Input Catalogue
(KIC). There are cases where the metallicity measurement is suspect,
and hence we also determine what happens if we do not use knowledge of
the metallicity.  We assume that the apparent magnitude of the star is
known exactly and that all the error lies in the parallax
measurement. However, since most KIC stars do not have accurate
parallaxes, we also determine what happens to the radius determination
if we do not know the parallax at all. The errors in the seismic
variables are roughly what is expected from a month of Kepler data.

\begin{figure*}
\epsscale{0.7}
\plotone{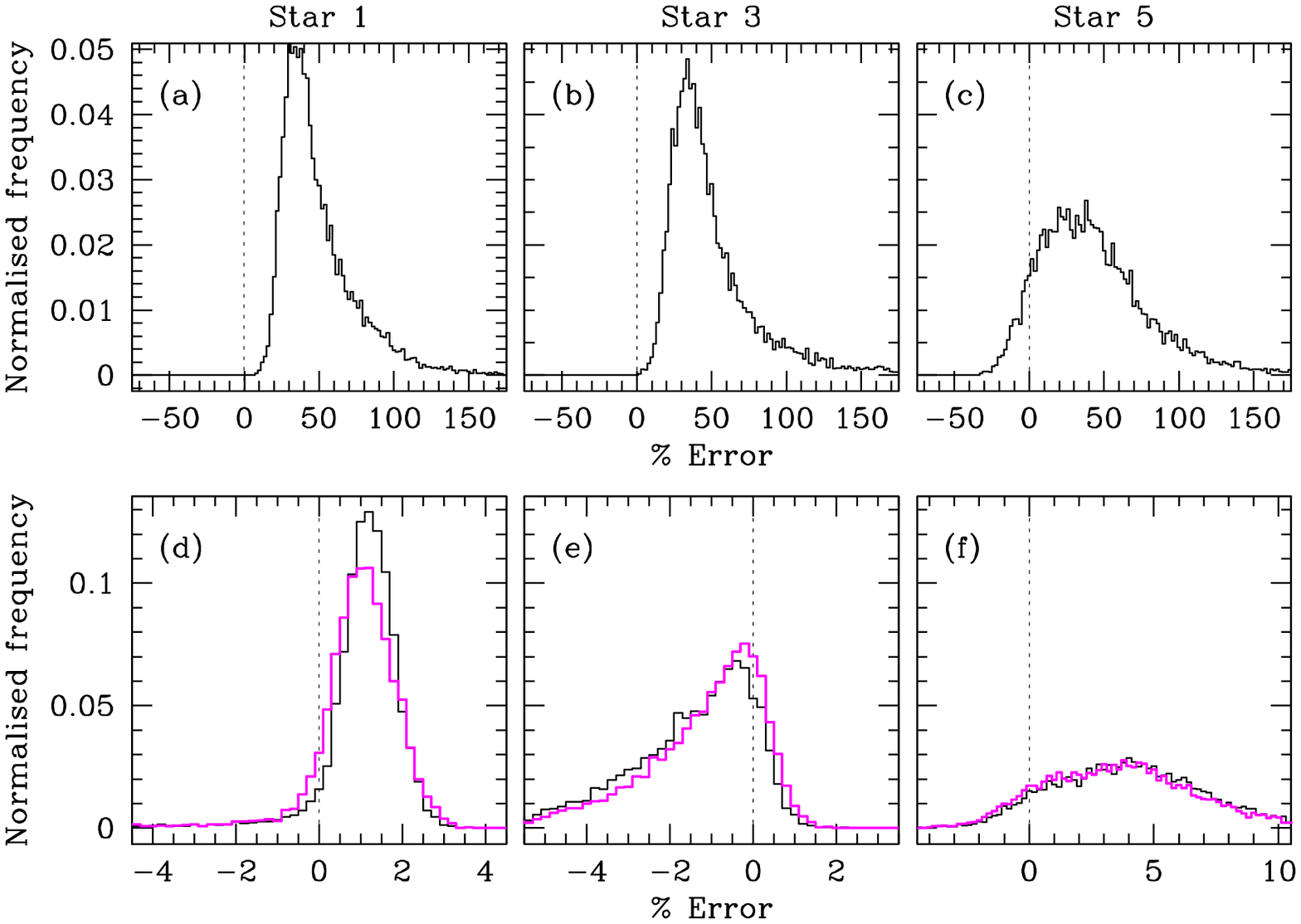}
\caption{The influence of seismic parameters on the radius
determination. Note the large change in the scale of the abscissa
(i.e., the change in the uncertainty in the estimated radius) once we
start using seismic observables. Panels (a), (b) and (c) show the
distribution function plotted as a function of the percentage error in
radius for Star~1, Star~3, and Star~5 respectively, when only the
non-seismic variables $T_{\rm eff}$, $Z/X$, apparent magnitude $V$ and
parallax $\pi$ are used. Error distribution Error~2 was used. Panels
(d), (e) and (f) show the results when seismic parameter $\Delta\nu$
is used (magenta lines, grey in the print version), as well as what happens when both $\Delta\nu$
and $\nu_{\rm max}$ (black lines).  We plot the distribution function
as a function of the percentage deviation from the actual result to
make it easier to gauge both systematic error (shift in the peak of
the distribution function away from zero) and statistical error (the
width of the function).}
\label{fig:var}
\smallskip
\epsscale{0.7}
\plotone{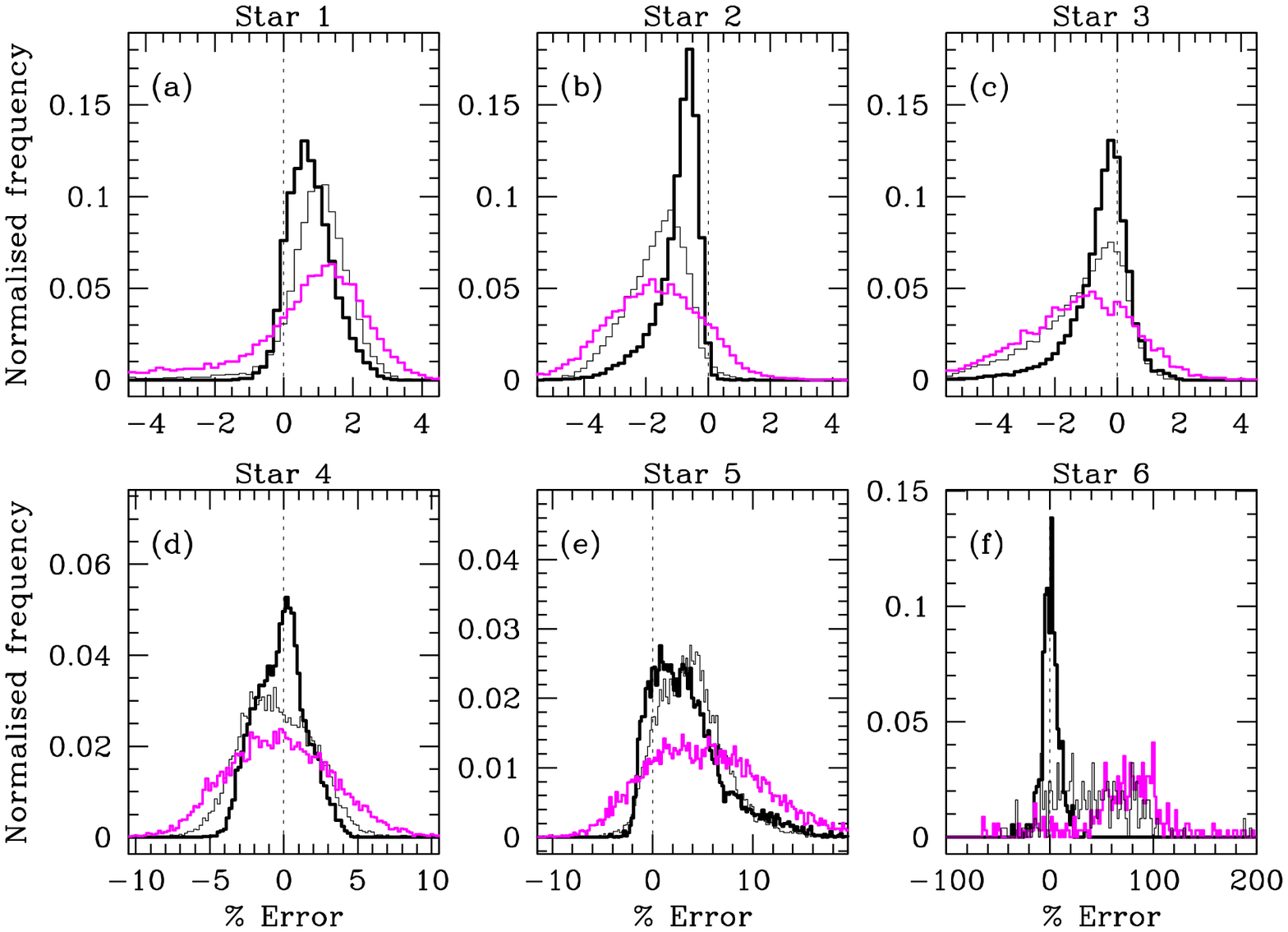}
\caption{The distribution function obtained for the 6 test stars
(Table~\ref{tab:teststars}) when the errors listed in
Table~\ref{tab:error} are used. The thick black line is for Error~1,
thin black line for Error~2 and magenta line (grey line in the print
version) for Error~3. Note that not
only do the distribution functions become wider, implying larger
statistical errors, when the errors in the inputs are increased the
systematic errors increase too.}
\label{fig:error}
\end{figure*}

Figure~\ref{fig:var} shows the distribution functions obtained for
three of the stars when we do not use seismic inputs, and for when we
do. The results are for Error~2 (cf. Table~\ref{tab:error}).  As is
clear, using only conventional variables ($T_{\rm eff}$, $Z/X$, $V$
and $\pi$) gives radius errors that can be 50\,\% or higher. When just
the large spacing ($\Delta\nu$) is added to the list of inputs, the
errors are brought down to less than 5\%. Adding the frequency of
maximum mode power ($\nu_{\rm max}$) helps a bit more, but the change
is not as spectacular.

Figure~\ref{fig:error} shows the effect of the three error
distributions (Errors~1, 2 and 3) on the resultant distribution
function. We remind the reader that the difference between the three
errors lies in differences in the assumed parallax, and errors
therein, as well as differences in the errors of the seismic
variables. As can be seen from the figure, as expected, the
distribution with the smallest errors does the best. Increasing errors
does not merely widen the distribution function (denoting larger
statistical errors in the solution), it also shifts the peak away from
the real solution (denoting systematic errors). The situation becomes
particularly bad for evolved stars. We fail dismally in the case of
the red giant (Star~6) to get any sensible result for Errors~2 and 3.
As we show later, estimates of effective temperature and absolute
magnitudes need to be good to be able to get radii of red-giants
accurately.  Given that many red giants in the Kepler sample are
expected to be bright enough to obtain individual frequencies even
from relatively short time series (mode amplitudes will be
significantly higher than in the less luminous main-sequence targets),
detailed modelling of the individual frequencies should give better
estimates of the radii of these stars.

\subsection{The effect of input errors}
\label{subsec:res}

Our radius-determination method uses a number of inputs, and the
errors in the inputs affect the final result. We determine the effect
of errors in the inputs by comparing the distribution functions
obtained with Error~2 with those obtained by reducing the error in
each input variable one by one.

\begin{figure}
\epsscale{0.45}
\plotone{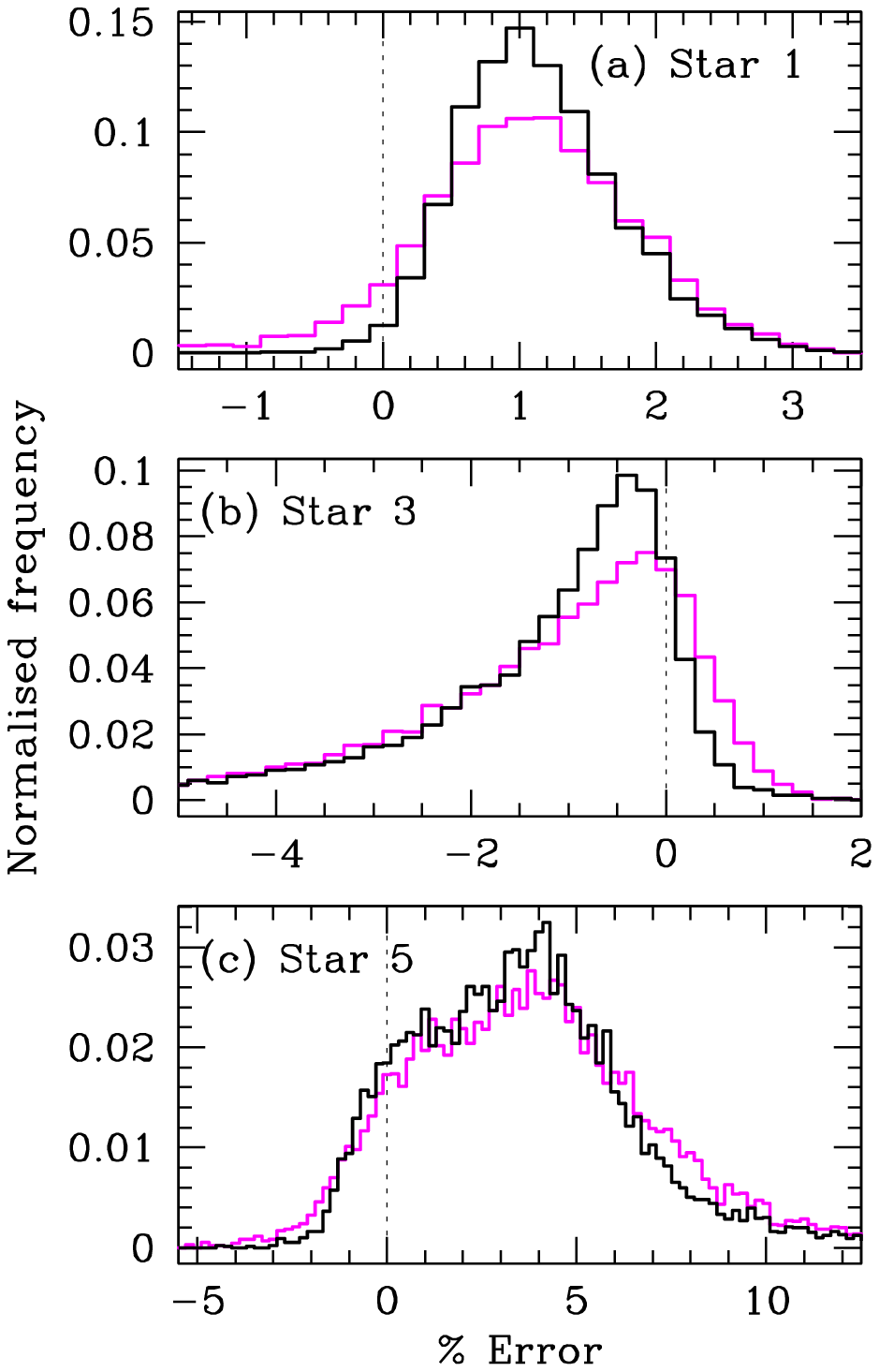}
\caption{The effect of error in $\Delta\nu$. We show the distribution
function obtained when the error in $\Delta\nu$ is reduced by a factor
of two (black line). The original distribution, obtained with
Error~2 (Table~\ref{tab:error}) is shown in magenta (grey).}
\label{fig:delnu}
%
\epsscale{0.45}
\plotone{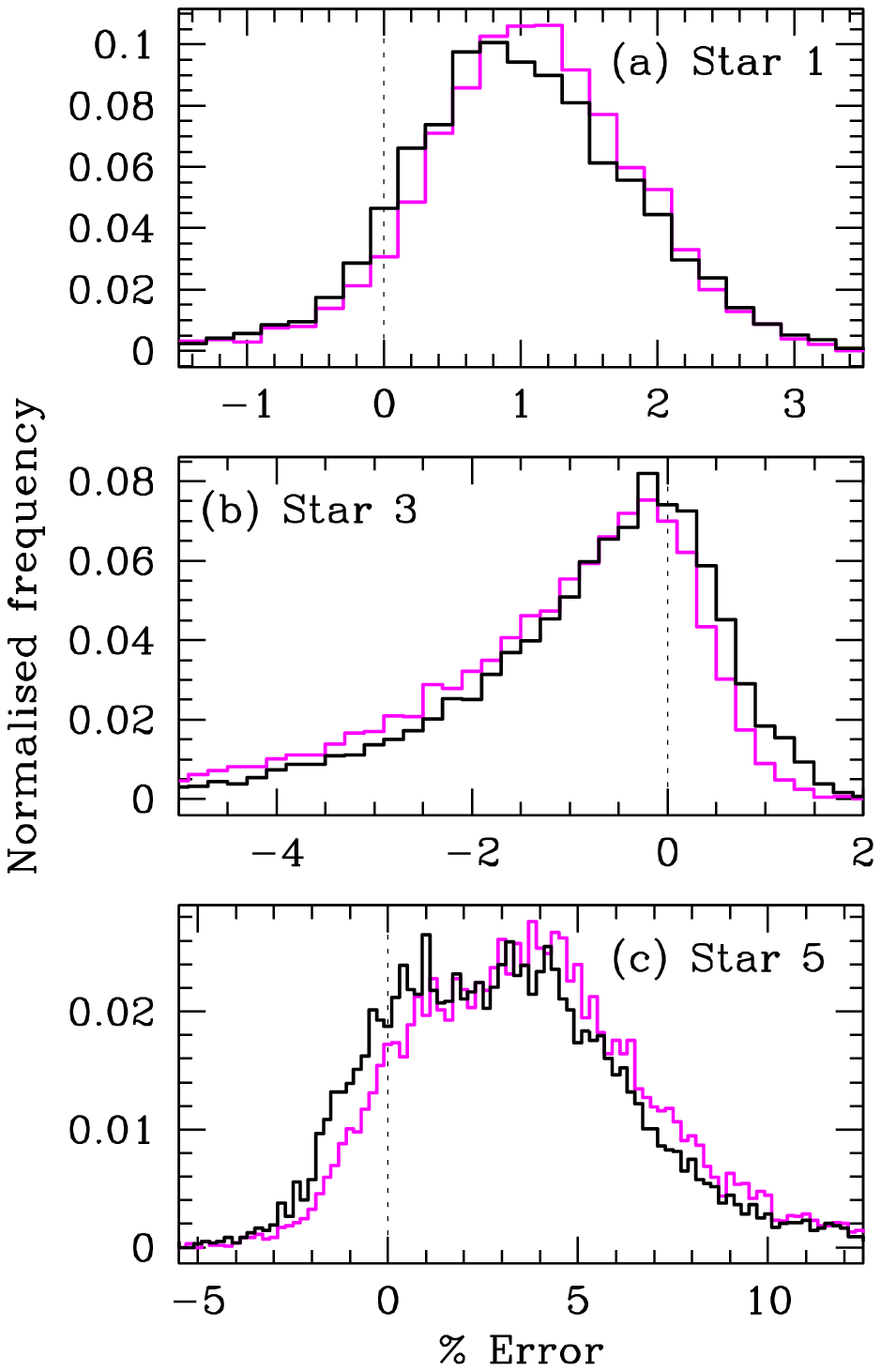}
\caption{The effect of the error in $\nu_{\rm max}$. We show the
distribution function obtained when the error in $\nu_{\rm max}$ is
reduced by a factor of two (black line). The original
distribution, obtained with Error~2 is shown in magenta (grey).}
\label{fig:numax}
\end{figure}

The effect on the distribution function of errors in the seismic
variables, $\Delta\nu$ and $\nu_{\rm max}$, is shown in
Figures~\ref{fig:delnu} and~\ref{fig:numax}, respectively. The error
in $\Delta\nu$ was reduced from $0.56\mu$Hz to $0.28\mu$Hz. As can be
seen, reducing the error in $\Delta\nu$ makes the distribution
functions narrower, and hence the results more precise. The effect is
marked for main-sequence and turn-off stars, but is minimal for
sub giants and red giants. As we shall show later, the precision in
$T_{\rm eff}$ and $(Z/X)$ is much more important for sub giants, while
the precision of the parallax is extremely important for
red giants. The effect of the precision of $\nu_{\rm max}$ is minimal,
unless the relative precision is of the same order as that of
$\Delta\nu$.

\begin{figure}
\epsscale{0.45}
\plotone{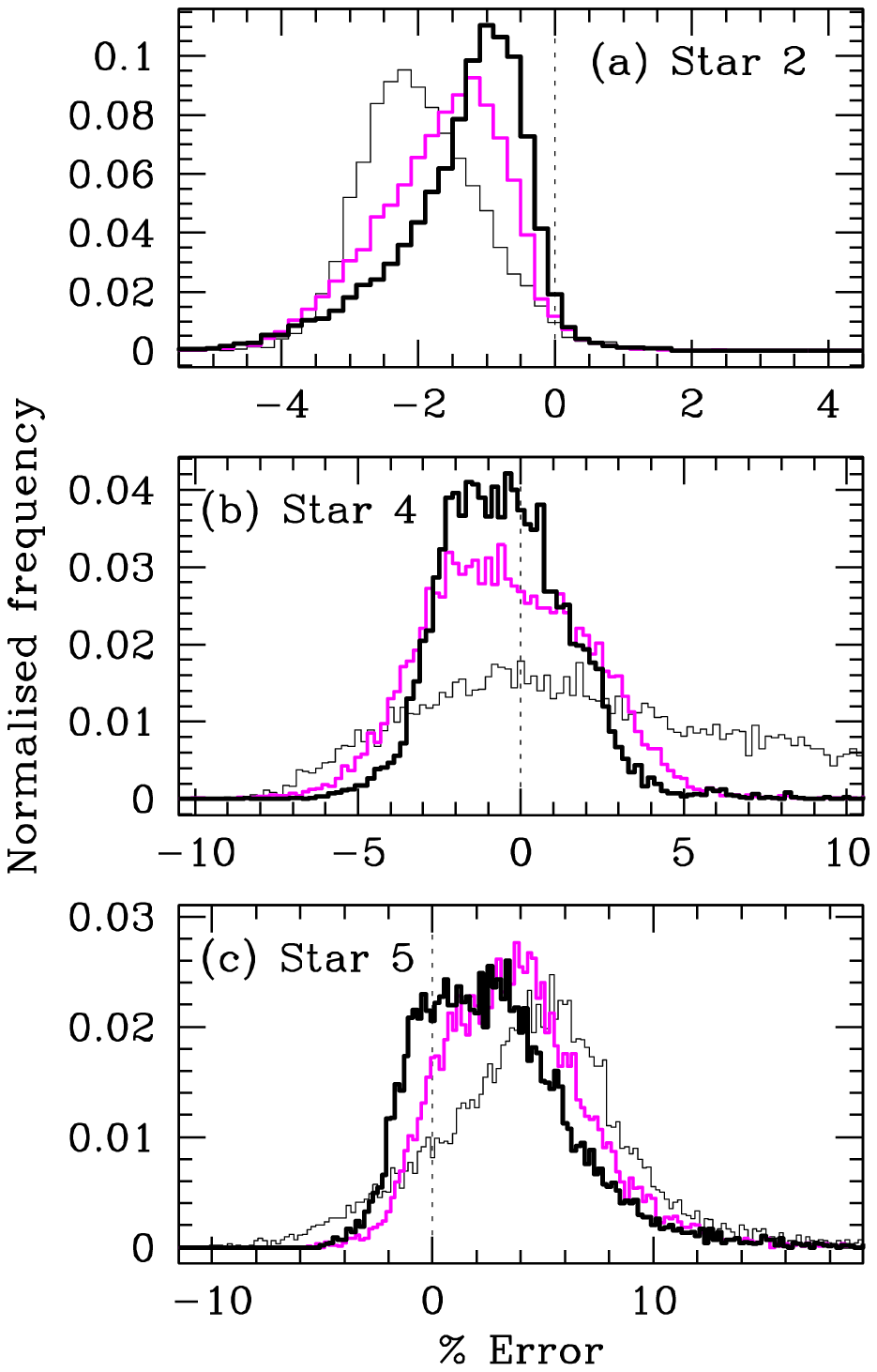}
\caption{The effect of the error in $T_{\rm eff}$. The distribution
functions are plotted as a function of the percentage error, including
when $T_{\rm eff}$ is not used at all (thin black line) and when the
error in $T_{\rm eff}$ is reduced by a factor of two (black lines)
compared to Error~2 (magenta, grey in the print version).}
\label{fig:thalf}
%
\epsscale{0.45}
\plotone{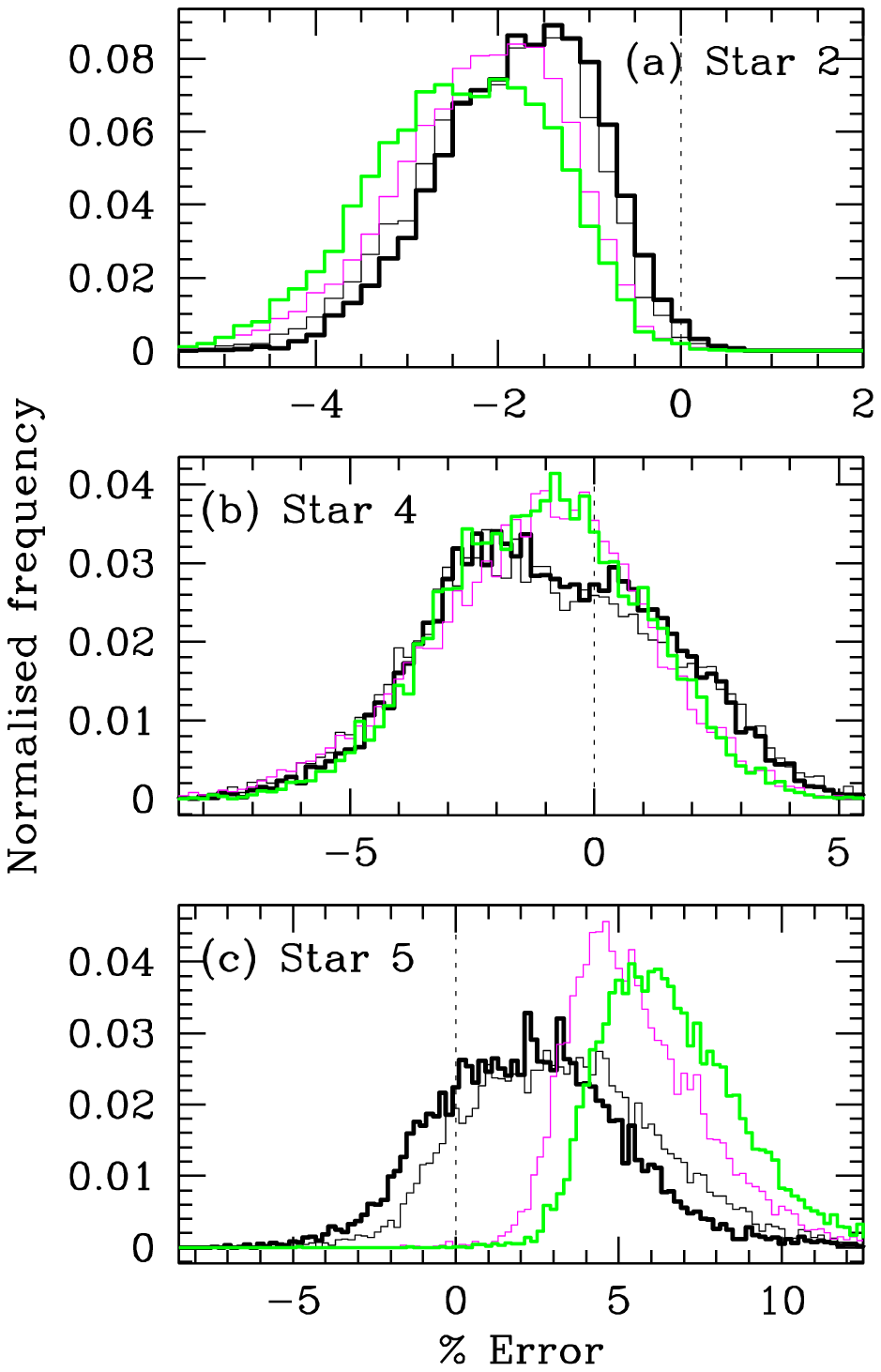}
\caption{The effect of the error in $Z/X$. The distribution functions
are plotted as a function of the percentage error in radius, including
when we disregard $Z/X$ altogether (magenta line, thinn grey in the
print version) compared to the
original result (thin black line). The distribution shown in the
thick black line is obtained when errors in $\log(Z/X)$ are reduced by
a factor of two compared to what was used for the dotted
distribution. The distribution in green (thick grey in the print version)
is what is obtained if we
assumed that the star has a solar metallicity, even if it does not.}
\label{fig:zpar}
\end{figure}
\begin{figure}
\epsscale{0.45}
\plotone{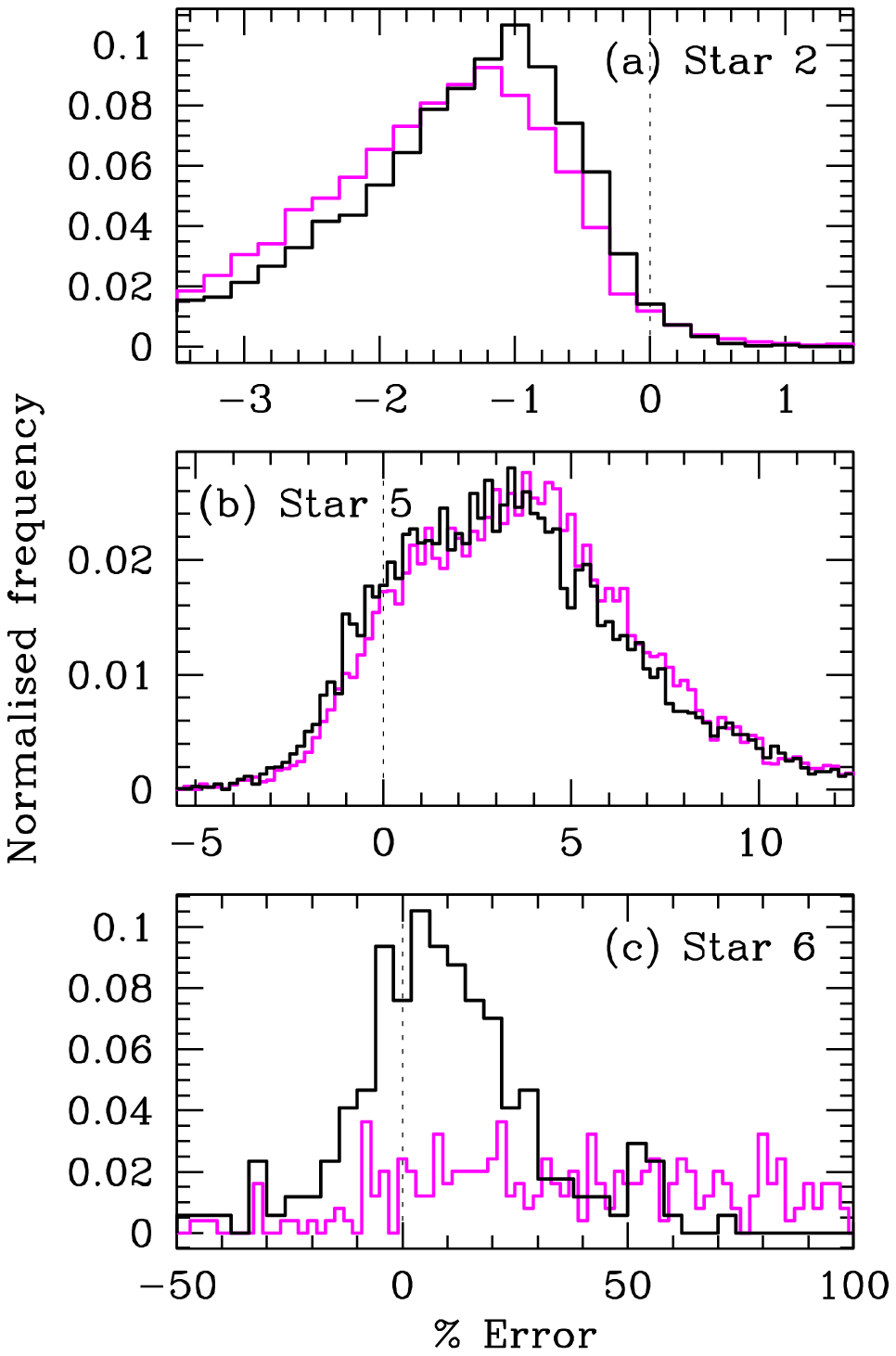}
\caption{The effect of the error in the parallax $\pi$. We show the
distribution function obtained when the error in $\pi$ is reduced by a
factor of two (black line). The original distribution, obtained
with Error~2, is shown in magenta (grey).}
\label{fig:parlow}
\smallskip
%
\epsscale{0.9}
\plotone{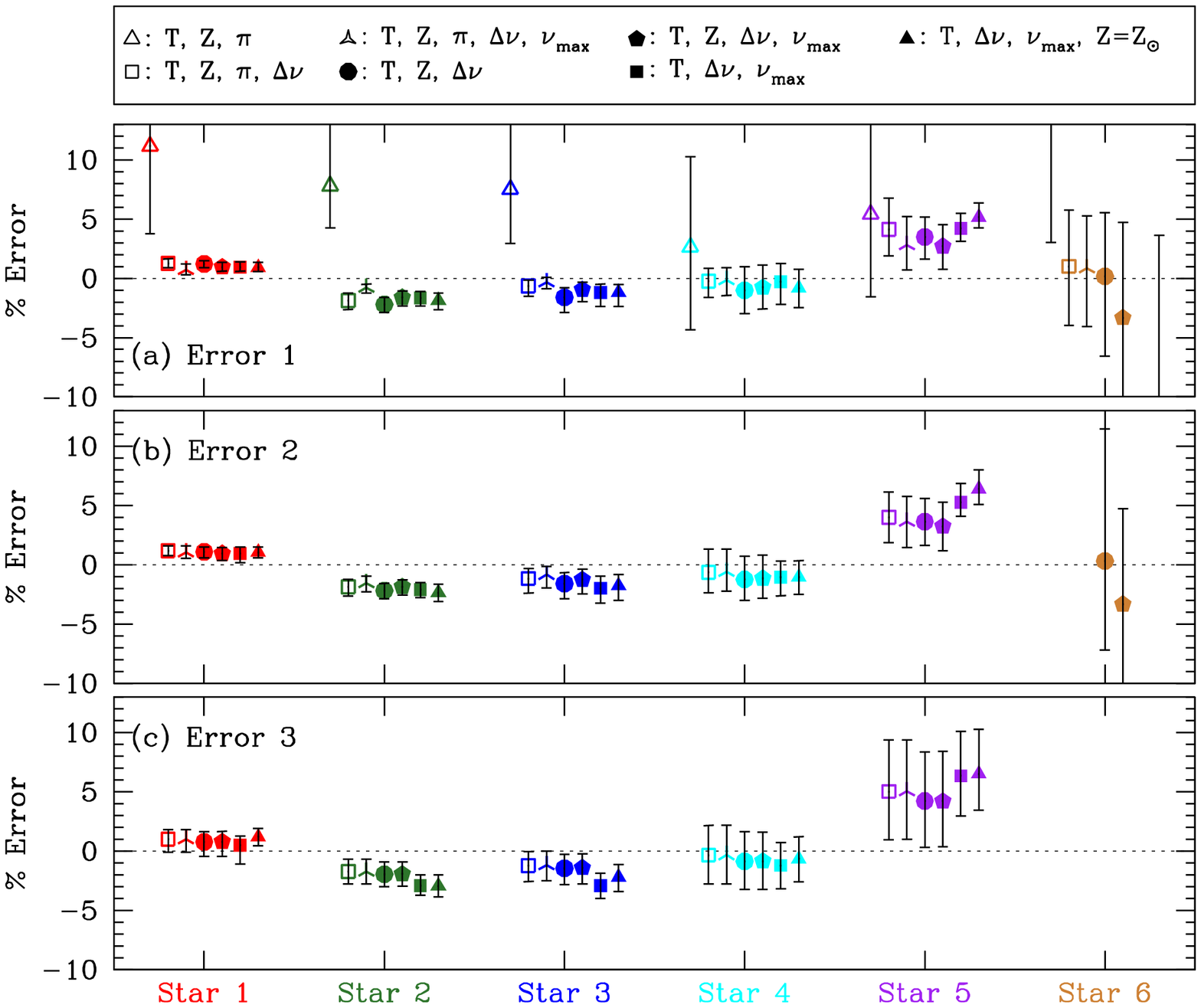}
\caption{The summary of our results with the six test models. We plot
the percentage error in each of the results. The panels are for the
three error distributions (Table~\ref{tab:error}). Each star is
indicated by a different colour and the symbols denote the combination
of inputs used.}
\label{fig:all}
\end{figure}

Figure~\ref{fig:thalf} examines the effect of $T_{\rm eff}$ on the
results. As can be seen, failure to use $T_{\rm eff}$ is not an
option. Decreasing errors in $T_{\rm eff}$ by a factor of two, down to
100\,K, increases both the accuracy and precision of the results and
additional improvement is obtained by decreasing $\sigma(T_{\rm eff})$
further. The r\^ole of $T_{\rm eff}$ becomes particularly important in
turn-off and post-main-sequence stars where a small change in
effective temperature is accompanied by a large change in radius.

The effect of metallicity is shown in Figure~\ref{fig:zpar}. We
obtained the distributions using Error~2 errors with and without
$Z/X$. We also tried reducing the error in $\log(Z/X)$ by a factor of
two. The last case we tried assumed that the star has solar
metallicity. This was prompted by the fact that the metallicity
distribution of solar neighbourhood stars is peaked at very nearly
solar metallicity. Our results show, however, that not using $Z/X$ at
all gives better results than assuming solar metallicity. The
difference is slight for main-sequence stars, but glaring for evolved
stars. We find that we cannot get a good radius measurement for
evolved stars without a good knowledge of their metallicity.

Figure~\ref{fig:parlow} shows the effect of errors in the
parallax. Reducing parallax errors does not affect the results of
main-sequence stars and most sub giants too much. It does make a very
large difference though in the case of red giants where the luminosity
is basically determined by the radius given their narrow $T_{\rm eff}$
range. In the case of main-sequence stars, the effect of even having a
parallax is small.  This is good news for initial analysis of Kepler
data, given that most of the stars in KIC do not have reliable
parallaxes (though, as noted previously, we expect eventually to get
parallax determinations from the Kepler data).

Figure~\ref{fig:all} summarises our results for the six test stars. We
show results for each of the three error distributions as well as the
results of using (or not using) certain inputs. We plot the deviation
of the inferred radius from the actual radius of the model; thus a
difference from zero indicates immediately the systematic error in the
result, while the error bars indicate the statistical errors. It can
be seen that by using seismic inputs we can determine the radius of
main sequence stars to within 1 to 2\,\%. The error in the sub giants
is somewhat higher, here about 5\,\%. We fail pretty badly in the case
of the red giant, except when the input errors are small (i.e., the
Error~1 case). The figure shows the non-seismic results too: except
for the case of Error~1, the error in those results always exceeds
15\,\%. The figure also shows that knowledge of $T_{\rm eff}$ and the
seismic variables is sufficient to allow a good determination of the
radii of main sequence stars. More information is required for sub
giants and red giants --- knowledge of $Z/X$ for sub giants, and both
$Z/X$ and luminosity (i.e., apparent magnitude and parallax) for red
giants. A good knowledge of the parallax is essential to get red giant
radii correctly.

\section{Possible sources of systematic errors}
\label{sec:sys}

There are two possible sources of systematic errors in the radius
determination, one is the so called ``surface term'' and the other is
the effect of the mixing length parameter.  We examine both these
sources in this section.

\begin{figure}
\epsscale{0.5}
\plotone{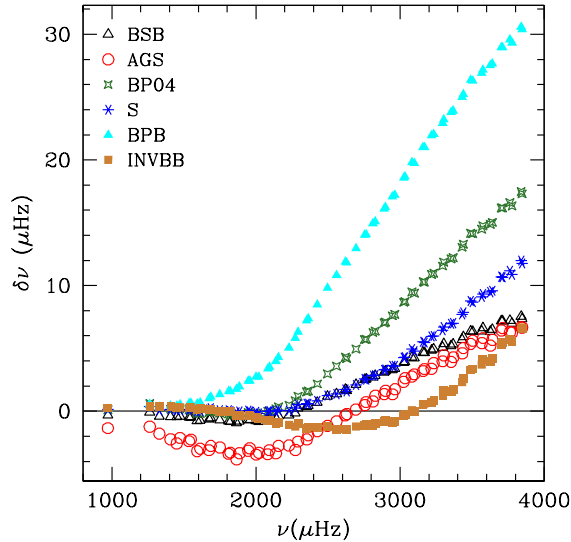}
\caption{The frequency differences between several solar models and
solar frequencies obtained from BiSON observations, plotted as a
function of the frequency of the mode. Differences for $\ell=0$-$3$
modes are shown. The predominant frequency-dependent trend in the
difference is a result of near-surface errors in modelling and is
commonly referred to as the ``surface term''. Note how the different
models have very different surface terms.}
\label{fig:surface}
\end{figure}

\subsection{The surface term}
\label{subsec:surf}

Stellar models do not represent the near-surface layers of stars very
well. Among the problems is the use of the mixing length approximation
to model convection. While this approximation works well in regions of
efficient convection, it does not in regions where convection is not
efficient, i.e., the near-surface layers. The structure of the
superadiabatic layer is incorrect (Nordlund \& Dravins 1990; Freytag
et al. 1996; Abbett et al. 1997, etc.). In addition, stellar models do
not include turbulent pressure that arises from convection, neither do
they include the contribution due to turbulent kinetic energy. All
these factors lead to errors in the models.  Near-surface errors of
this kind lead to a frequency-dependent error in the frequencies (see
e.g., Christensen-Dalsgaard \& Berthomieu 1991) usually referred to as
the ``surface term''.  Differences in near-surface opacities also
contribute.  To the contribution due to improper modelling is added
the fact that frequencies of models are calculated assuming that they
are adiabatic, another surface term since that is where
adiabaticity breaks down.

The surface term for various published solar models can be seen in
Figure~\ref{fig:surface}.  The models are listed in
Table~\ref{tab:solar}.  This figure shows frequency differences for
modes with $\ell=0,1,2$ and $3$ between the models and the solar
frequencies in the aforementioned BiSON-1 set, plotted as a function
of frequency. As can be seen, the frequency differences are
predominantly a function of frequency; furthermore, different models
have different surface terms.  Methods have been developed by
helioseismologists to deal with the surface term in solar data (e.g.,
Dziembowski et al.~1990). Kjeldsen et al. (2008) have also recently
proposed an empirical method for constraining the surface term using
asteroseismic data \textit{only when} estimates of individual mode
frequencies are available. However, this is not an option for those
Kepler Survey Phase targets where we can only extract an estimate of
the average large frequency spacing, and cannot extract robust
estimates of individual frequencies (covering several radial
orders). Thus we might reasonably expect the radius determination in
these stars to be affected, because of the effect of the surface term.

\begin{deluxetable}{llc}
\tablecaption{The effect of the ``surface term'' --- results for
  various solar models}
\tablehead{
\colhead{Model}&\colhead{Comments}&\colhead{Inferred Radius ($R_\odot$)}}
\startdata
BSB & Model BSB(GS98) of Bahcall et al. (2006)& $1.0023_{-0.0166}^{+0.0135}$ \\ 
\noalign{\smallskip}
AGS & Model BSB(AGS05) of Bahcall et al. (2006)& $1.0018_{-0.0169}^{+0.0133}$ \\ 
\noalign{\smallskip}
BP04 & Model BP04 of Bahcall et al. (2005) & $0.9997_{-0.0180}^{+0.0135}$ \\ 
\noalign{\smallskip}
S & Model S of Christensen-Dalsgaard et al. (1996) & $1.0014_{-0.0172}^{+0.0132}$ \\ 
\noalign{\smallskip}
BPB & Model STD of Basu et al. (2000)  & $0.9961_{-0.0188}^{+0.0139}$ \\ 
\noalign{\smallskip}
INVBB & Seismic model of Antia (1996) & $1.0023_{-0.0166}^{+0.0135}$ \\ 
\enddata
\label{tab:solar}
\end{deluxetable}

To examine how the surface term can affect radius determination using
the large frequency spacings, we calculated the average large
spacings for each of the solar models shown in
Fig.~\ref{fig:surface} and tried to determine their radii with the
same error-distribution as Error~2. Although the large spacings of
the models do not include all effects of the surface term, since they
too are mixing-length models, they are nonetheless sufficiently
different from the calibration models used in the radius search
method. The results are listed in Table~\ref{tab:solar}. As can be
seen, at least for these models, the differing surface terms do not
affect the results much, and the systematic error is much less than
1\,\%, and smaller than the statistical error. We do not therefore
believe that we need to be worried about the surface term when dealing
with data from the Kepler survey phase.  The results are, however,
better than the BiSON results discussed in Section~\ref{sec:method}
and perhaps we should consider those results as an indication of what
systematic errors the surface term introduces.

\subsection{The mixing length}
\label{subsec:mlt}

The radius of a stellar model depends on the mixing length parameter
used to construct the model. Unfortunately, there is no method to
determine what the mixing length parameter should be: usually the
mixing length parameter needed to reproduce the solar radius at the
solar age is used for models of other stars too, and it is quite
possible that this is not the correct thing to do. In fact simulations
of convection suggest that the mixing length parameter varies across
the HR diagram (e.g. Ludwig et al. 1999, 2002; Robinson et
al. 2004). All the calibration models here were, of course,
constructed with the solar value of the mixing length, and hence it is
possible that we will get erroneous results for some stars.  To judge
how badly we could fail, we test out our method using models
constructed with mixing lengths that are different from the one used
to construct our calibration models.

\begin{deluxetable}{lccccccl}
\tablecaption{Characteristics of the models with different mixing length parameters}
\tablehead{
\colhead{\ } & \colhead{$Z/X$} & \colhead{$T_{\rm eff}$} & \colhead{$M_{\rm V}$} & 
\colhead{$\Delta\nu$} & \colhead{$\nu_{\rm max}$} & \colhead{Radius} & \colhead{Comments} \\
\colhead{\ }     & \colhead{\ }  &  \colhead{(K)} & \colhead{\ }  &  \colhead{($\mu$Hz)}
 & \colhead{($\mu$Hz)} & \colhead{(R$_\odot$)}& \colhead{\ } }
\startdata
\multispan8{\hfill$\alpha=1.2H_p$\hfill}\\
Test 1-1 &  0.024 &  5360 & 5.11 & 123.56             & 2902 & 1.06 & Main sequence star\\
Test 1-2 &  0.024 &  5386 & 4.54 & {\phantom{1}}85.40 & 1769 & 1.36 & Near turn-off \\
Test 1-3 &  0.024 &  5219 & 4.30 & {\phantom{1}}63.77 & 1218 & 1.65 & Sub giant \\
\multispan8{\hfill$\alpha=2.4H_p$\hfill}\\
Test 2-1 &  0.024 &  6059 & 4.70 & 146.53             & 3426 & 0.95 & Main sequence star\\
Test 2-2 &  0.024 &  6128 & 4.22 & 109.29             & 2307 & 1.15 & Near turn-off \\
Test 2-3 &  0.024 &  5613 & 3.87 & {\phantom{1}}62.60 & 1145 & 1.67 & Sub giant\\
\enddata
\label{tab:mlt}
\end{deluxetable}

\begin{figure*}
\epsscale{0.7}
\plotone{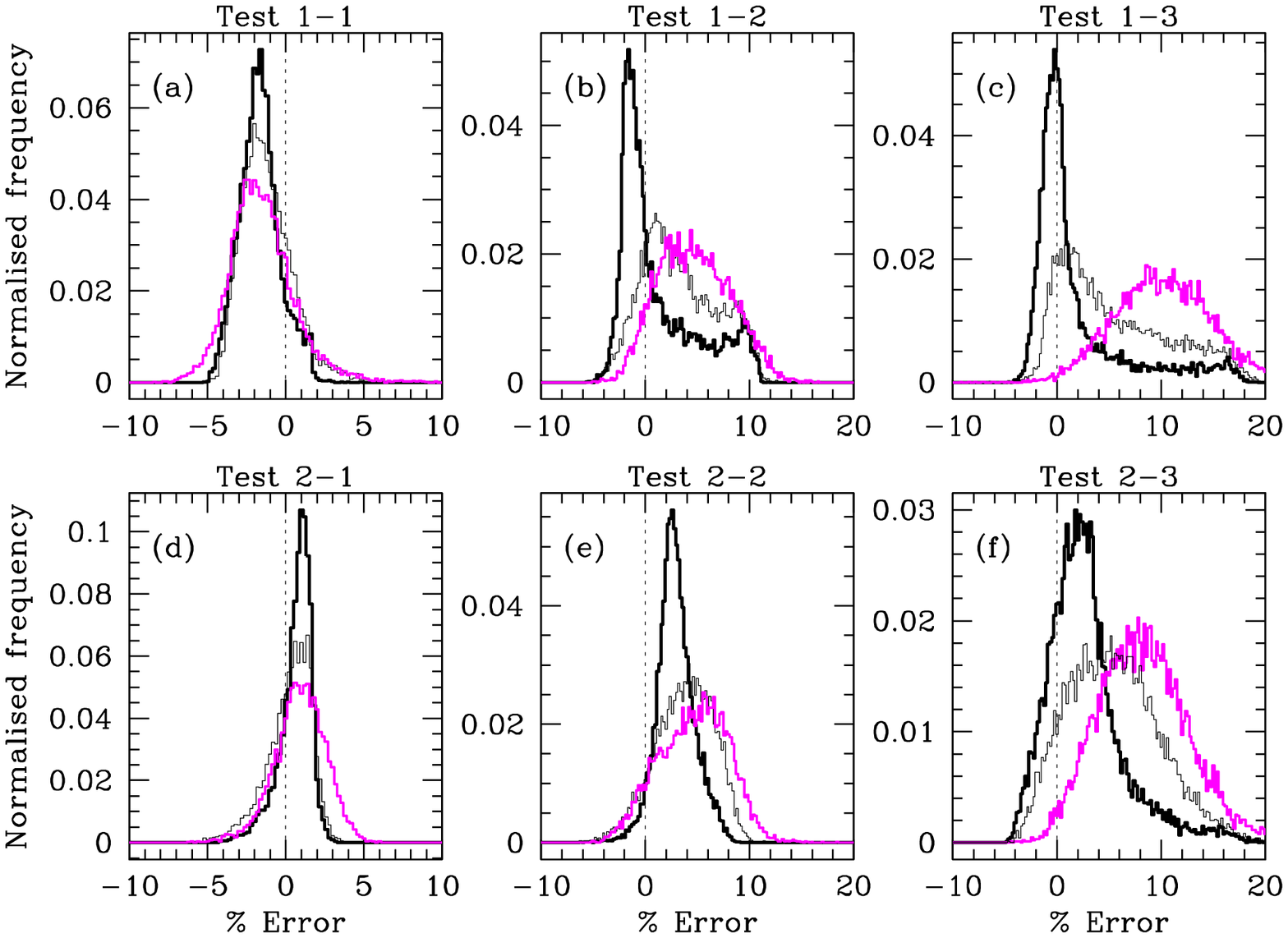}
\caption{Distribution functions for the the six models that have
different mixing length parameters than the calibration models. The
models are summarised in Table~\ref{tab:mlt}.  The thick black, thin
black and magenta (grey) lines correspond to Errors~1-3 (Table~\ref{tab:error})
respectively.}
\label{fig:mlt}
%
\smallskip
\epsscale{0.9}
\plotone{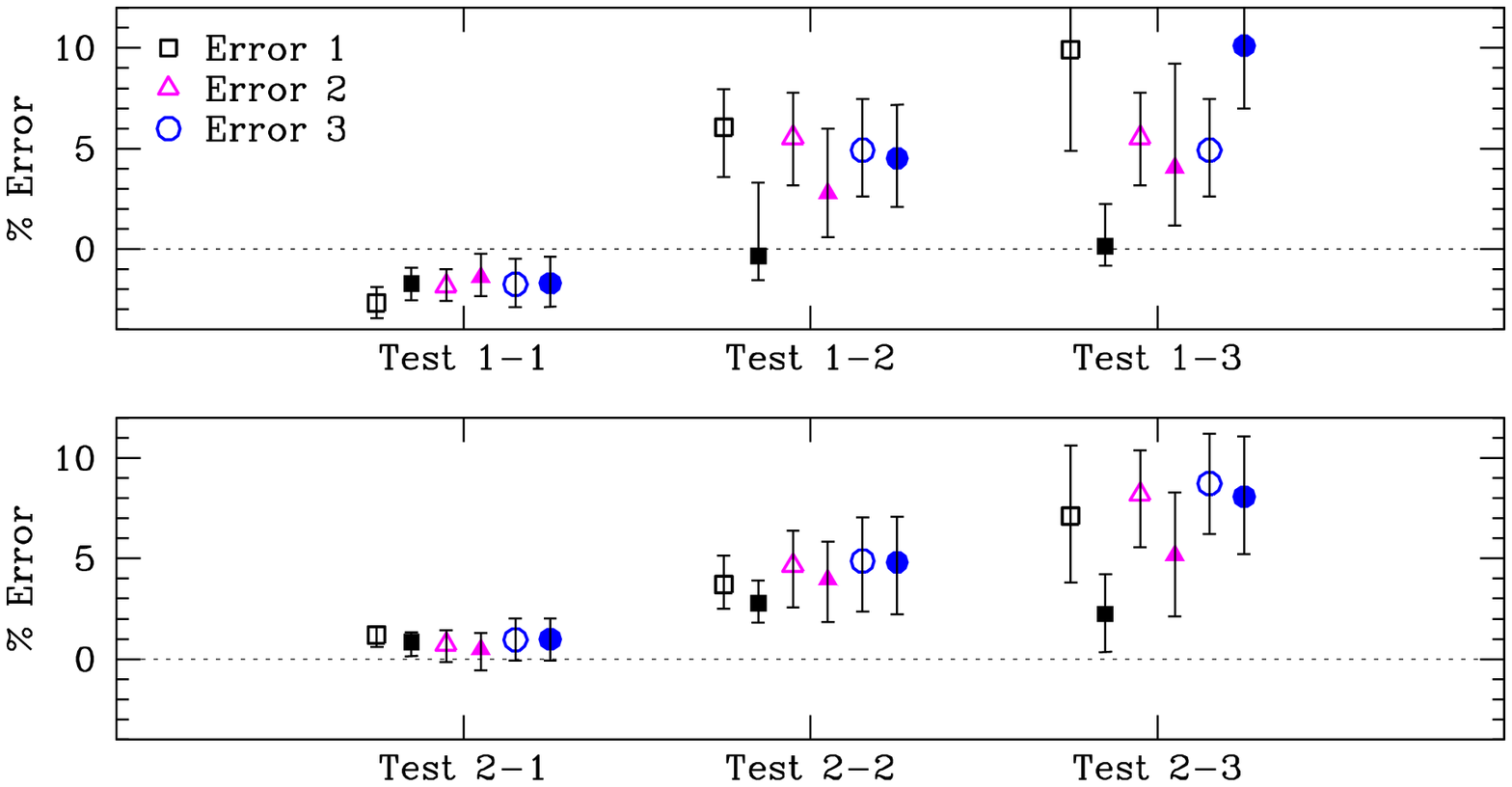}
\caption{The summary of our results for the six models with different
mixing length parameters. The black, red and blue points are the
results for the different error distributions, Errors~1-3 respectively.
The filled symbols are for the results when all inputs (i.e., $T_{\rm
eff}$, $Z/X$, $\pi$, $\Delta\nu$ and $\nu_{\rm max}$) are used, the
open symbols are the result of not using $\nu_{\rm max}$.}
\label{fig:mlt_all}
\end{figure*}

We use a set of six models constructed with two different mixing
lengths. The models are listed in Table~\ref{tab:mlt}.  For each
mixing length we have one main sequence, one turn-off and one
sub-giant model. Given our failure with red giants in
Section~\ref{sec:test}, we do not try any here. We use the same error
distributions as before, i.e., the ones in Table~\ref{tab:error}. The
distribution functions obtained for the six models are shown in
Figure~\ref{fig:mlt}. We used all inputs ($T_{\rm eff}$, $Z/X$, $V$,
$\pi$, $\Delta\nu$ and $\nu_{\rm max}$) to obtain the distribution
functions. We can see that there are indeed shifts in the distribution
function away from the true value, when a mismatch of the mixing
lengths is present. The shifts are particularly bad for the 
evolved stars when the errors in the input parameters are large. In the
case of main-sequence models, the error is no larger than that for
the models used in Section~\ref{sec:test}.  The results in terms of the
radii obtained are summarised in Figure~\ref{fig:mlt_all} where we
also show what happens if we do not use $\nu_{\rm max}$. As we can see
from the figures, the difference between the inferred and the true
radius increases as the star evolves. The difference appears to be at
the 1 to 2$\sigma$ level.  The largest error we get is about 10\,\%, which
is less than what we would get if we did not have seismic parameters,
but is nevertheless quite large. We also see that the extra information
in $\nu_{\rm max}$ helps in most cases. Thus there is a distinct
possibility that our results obtained with Kepler data for evolved
stars will be affected. It will help if the errors in the inputs are
as small as possible.

\section{Blind tests}
\label{sec:hh}

The stellar models that we tested in Section~\ref{sec:test} were very
similar to the calibration models that were used to determine the
distribution functions. We did use different models in
Sections~\ref{subsec:surf} and \ref{subsec:mlt}, however, as for the
work described in Section~\ref{sec:test}, our inputs were exact,
without errors. The nominal errors were used only to generate the
models that were needed to determine the distribution function.

As a final check on our methods, we therefore performed a proper
``all-up'' blind test of the technique involving models constructed
with different codes (including different mixing length parameters),
errors on the inputs, and a surface term included. To this end, the
first author of this paper (SB) was given a series of input parameters
by the one of the other co-authors (WJC) to determine the radius of
the models. The exact parameters and radii were disclosed later. The
actual characteristics of the three asteroFLAG models we used (three
of the so-called ``asteroFLAG cats'') are listed in
Table~\ref{tab:blindchar}; while the inputs provided for the test are
listed in Table~\ref{tab:blindinp}. We now go on to describe how these
input data were generated by WJC.

First, we take the non-seismic inputs, i.e., $T_{\rm eff}$,
$\log(Z/X)$, and $\pi$. To the actual $T_{\rm eff}$ and $\log(Z/X)$ we
added a random Gaussian deviate that had been multiplied by the
uncertainty expected on similar data in the KIC. True values of the
parallax, $\pi$, were calculated for two Kepler apparent magnitudes,
$V=9$ and $V=10$. As noted previously, an Hipparcos-like uncertainty
was then assigned to each parallax, and the true value of the parallax
perturbed (in the manner described for $T_{\rm eff}$ and $\log(Z/X)$)
before the data were passed to SB.

The seismic inputs came from tests conducted on artificial oscillation
power spectra of the three cats, which were designed to mimic
observations expected in the Survey Phase. These spectra were made
with the asteroFLAG simulator. Full details may be found in Chaplin et
al. (2008c) [see also Stello et al. 2009]. Here, we note that the
stellar models were generated using the Aarhus stellar evolution code,
ASTEC (Christensen-Dalsgaard 2008a), while the oscillation frequencies
were calculated using the adiabatic pulsation code ADIPLS
(Christensen- Dalsgaard 2008b). The oscillation spectra of the cats
differed from those used in Stello et al. (2009) in two important
respects. First, the underlying oscillation amplitudes and damping
rates were modified to more accurately reflect expected levels for the
Kepler observations (e.g., based on experience accumulated from
analysing the CoRoT data). Second, artificial spectra were made to
simulate those expected from observations lasting one month (i.e., the
Survey Phase length), not the 3.5\,yr used in Stello et al.

\begin{deluxetable}{lcccccc}
\tablecaption{Characteristics of asteroFLAG blind-test models}
\tablehead{
\colhead{Name} & \colhead{Mass} & \colhead{Radius} & \colhead{Age} & \colhead{$\alpha$} &$T_{\rm eff}$& $\log(Z/X)$ \\
               &  \colhead{($M_\odot$)}&\colhead{($R_\odot$)}&\colhead{(Gyr)}&\colhead{\ }& \colhead{(K)} & \colhead{\ } }
\startdata
Katrina & 0.70 & 0.63 & 1.000 & 3.05  & 4350.03 &  $-1.67$ \\
Boris   & 1.00 & 1.00 & 4.600 & 1.99  & 5777.00 & $-1.61$ \\
Pancho  & 1.40 & 1.75 & 1.995 & 2.03  & 6351.57 & $-1.39$ \\
\enddata
\label{tab:blindchar}
\end{deluxetable}

\begin{deluxetable}{lccccccc}
\tablecaption{Inputs for the blind test on three of the asteroFLAG cats}
\tablehead{
\colhead{Name} & \colhead{$T_{\rm eff}$} & \colhead{$\log(Z/X)$} & \colhead{$V$} & \colhead{$\pi$} &
\colhead{$\Delta\nu$} & \colhead{$\nu_{\rm max}$}\\
\colhead{\ } & \colhead{(K)} & \colhead{\ } & \colhead{(mag)} & \colhead{(mas)} &
\colhead{($\mu$Hz)} & \colhead{($\mu$Hz)}}
\startdata
Katrina & $4618\pm200$ & $-1.42\pm0.2$& 9 & $46.8\pm1.2$ & $227.33\pm 1.89$ & $7209.1\pm 285$ \\
\noalign{\smallskip}
Boris & $5676\pm 200$ & $-1.31\pm0.2$ & 9 & $12.5\pm 1.2$ & $136.10\pm 0.30$ & $3165.4\pm 110$ \\
      &                 &                &10 & $8.8\pm 1.7$  & $136.29\pm 0.60$ & $3138.3\pm 200$ \\
\noalign{\smallskip}
Pancho & $6251\pm200$ & $-1.09\pm0.2$ & 9 & $8.9\pm 1.2$& $69.51\pm 0.34$ & $1684.8\pm 80$\\
       &                &               &10&  $3.8\pm 1.7$& $69.67\pm 0.56$ & $1671.2\pm 125$\\
\enddata
\label{tab:blindinp}
\end{deluxetable}

Many independent realisations of the oscillation power spectra were
made at each $V$, and the spectra analysed with the ``Octave''
(Birmingham-Sheffield Hallam) Kepler data-analysis pipeline to extract
the seismic inputs. This pipeline has been developed as part of the
asteroFLAG programme, and will be described in detail elsewhere
(Hekker et al., in preparation). We show only inputs for Katrina at
$V=9$ because it was hard to extract a robust estimate of the seismic
inputs from her artificial oscillation power spectra at $V=10$ (it has
the weakest modes of the three cats).

\begin{figure}
\epsscale{0.55}
\plotone{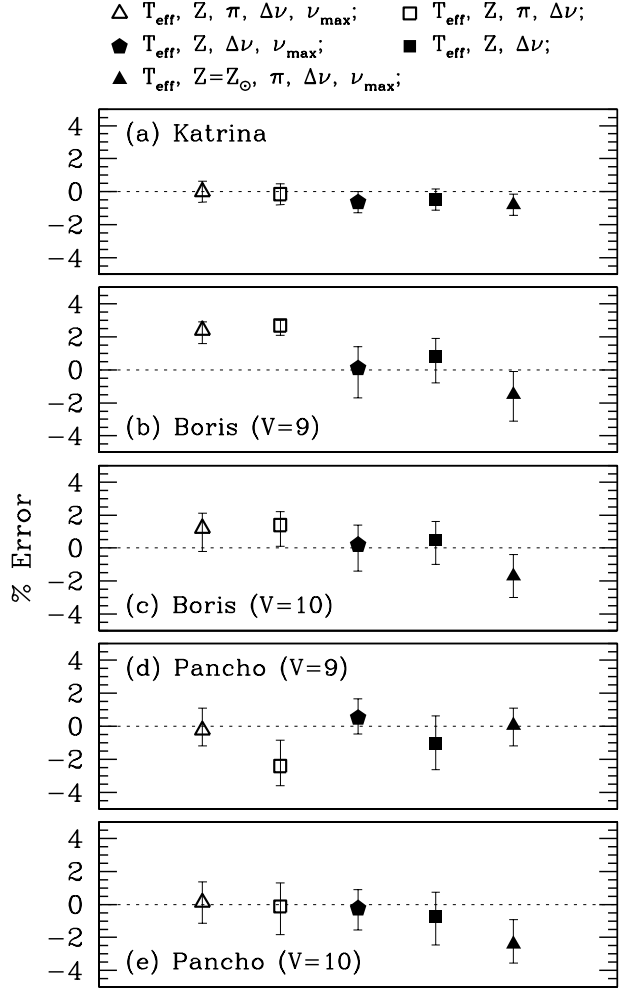}
\caption{The results of the blind tests plotted as the deviation from the
true value. Note that in all cases we are able to reproduce the results
to within a few percent. The different symbols denote the different
combinations of inputs used to obtain the results. The symbols are
explained at the top of the figure. The non-seismic inference (Case
(1) in Table~\ref{tab:blindres}) has not been plotted.}
\label{fig:blind}
\end{figure}

The results of our tests are listed in Table~\ref{tab:blindres}, and
shown graphically in Figure~\ref{fig:blind}. As can be seen from the
figure, we have succeeded in inferring the radii of the cats to within
a few percent. We are therefore confident that the Yale-Birmingham
method can be used to obtain reliable stellar radii using seismic data
from Kepler. The best results are seen for the low-mass main-sequence
star, as expected from what we have seen in Section~\ref{sec:test}.

\begin{table*}
\caption{Results of the blind test on three of the asteroFLAG cats}
\medskip
\begin{center}
\begin{tabular}{lcccccccc}
\noalign{\smallskip}
\hline
\hline
{Name} & {Radius} & {V} & \multicolumn{6}{c}{Inferred Radius}\\
 &(Exact) &  & (1) & (2) & (3) & (4) & (5) & (6)\\
 & ($R_\odot$)&  & \multicolumn{6}{c}{($R_\odot$)}\\
\noalign{\smallskip}
\hline
\noalign{\smallskip}
Katrina & 0.63 & 9 & $0.667 _{-0.024}^{+0.042} $ &  $0.630 _{-0.004}^{+0.004} $ &  $0.629 _{-0.004}^{+0.004} $ &  $0.626 _{-0.004}^{+0.004} $ &  $0.627 _{-0.004} ^{+0.004}$ &  $0.625 _{-0.004}^{+0.004} $ \\
\noalign{\smallskip}
Boris   & 1.00 & 9 & $1.292 _{-0.092}^{+0.112} $ &  $1.023 _{-0.008}^{+0.005} $ &  $1.027 _{-0.006}^{+0.003} $ &  $1.001 _{-0.018}^{+0.013} $ &  $1.008 _{-0.016} ^{+0.011}$ &  $0.985 _{-0.016}^{+0.014} $ \\
        &      &10 & $1.306 _{-0.110}^{+0.170} $ &  $1.012 _{-0.014}^{+0.009} $ &  $1.014 _{-0.013}^{+0.008} $ &  $1.002 _{-0.016}^{+0.012} $ &  $1.005 _{-0.015} ^{+0.011}$ &  $0.983 _{-0.013}^{+0.013} $ \\
\noalign{\smallskip}
Pancho  & 1.75 & 9 & $1.493 _{-0.092}^{+0.142} $ &  $1.746 _{-0.017}^{+0.023} $ &  $1.708 _{-0.021}^{+0.027} $ &  $1.759 _{-0.017}^{+0.020} $ &  $1.732 _{-0.028} ^{+0.030}$ &  $1.752 _{-0.022}^{+0.018} $ \\
        &      &10 & $4.610 _{-0.351}^{+0.606} $ &  $1.752 _{-0.021}^{+0.022} $ &  $1.748 _{-0.030}^{+0.025} $ &  $1.746 _{-0.023}^{+0.020} $ &  $1.737 _{-0.030} ^{+0.026}$ &  $1.708 _{-0.020}^{+0.026} $ \\
\noalign{\smallskip}
\hline
\noalign{\smallskip}
\end{tabular}

(1) Using $T_{\rm eff},\;Z,\;\pi$; (2) Using $T_{\rm eff},\;Z,\;\pi,\;\Delta\nu,\;\nu_{\rm max}$; (3) Using $T_{\rm eff},\;Z,\;\pi,\;\Delta\nu$\\
(4) Using $T_{\rm eff},\;Z,\;\Delta\nu,\;\nu_{\rm max}$; (5) Using $T_{\rm eff},\;Z,\;\Delta\nu$;
(6) Using $T_{\rm eff},\;Z=Z_\odot,\;\Delta\nu,\;\nu_{\rm max}$\
\end{center}
\label{tab:blindres}
\end{table*}

Although we used the same cats as those in used Stello et al. (2009),
we cannot directly compare our results with those since the errors we
assumed were different (higher in most cases). However, despite that,
our results compare favourably, as can be seen if we compare our
Table~\ref{tab:blindres} with Table~2 of Stello et al.~(2009). In
order to facilitate a direct comparison, we put their Katrina input
K1, Boris input B1 and and Pancho input P1 through our pipeline and
obtained radii of $0.625^{+0.003}_{-0.004}R_\odot$ for Katrina,
$1.020^{+0.007}_{-0.010}R_\odot$ for Boris and
$1.717^{+0.030}_{-0.036}R_\odot$ for Pancho. The results in Stello et
al.~(2009) range from $0.602$-$0.682R_\odot$ for Katrina,
$0.977$-$1.02R_\odot$ for Boris and $1.676$-$1.746R_\odot$ for Pancho.
Thus we compare very well with the other methods. It should be noted
that the exercise in Stello et al.~(2009) did not include $\nu_{\rm
max}$ and hence we did not use it either to determine the radius from
inputs K1, B1 and P1.

\section{Conclusions}
\label{sec:disc}

We have described a method to determine stellar radius using a mixture
of conventional and seismic variables. We use 
$\Delta\nu$, the average large frequency spacing, and $\nu_{\rm
max}$, the frequency of maximum mode power, as the seismic inputs; while
effective temperature, metallicity, the apparent visual magnitude and
parallaxes are the ``conventional'' inputs. We have also performed an error
analysis to determine which parameters play the most important r\^ole
in determining the radius using our method.

Our main conclusions are as follows:

\begin{itemize}

\item[--] We find that the mere presence of seismic data reduces
errors in the inferred radii to a few percent, with main-sequence
stars faring better than more evolved solar-type stars.

\item[--] We find that for main-sequence stars, a knowledge of the
parallax is not important. As long as the effective temperature and
the seismic variables are well constrained observationally, we can get
results to an accuracy of better than a few percent.  Metallicity does
not make much difference either for main sequence stars.  This is good
news as far as Kepler targets are concerned since the parallaxes and
metallicities in the Kepler Input Catalogue are known to be suspect,
when considered on a star-by-star basis.  Insensitivity to the
metallicity does not hold for all radius-determination methods, as can
be seen in Stello et al.~(2009).

\item[--] The situation is different for sub giants, and we find that
for them we need a good estimate of the metallicity (along with the
effective temperature and the seismic parameters) to get a reasonably
accurate result. However, we should emphasise that because of the
assumed availability of seismic data, we can still infer the radius of
these stars to much higher precision than would be possible without
those seismic data, no matter how poorly we know the metallicity of
the star.

\item[--] Robust estimation of the radii of red giants is extremely
difficult. The narrow temperature range they occupy, and the sensitive
dependence of the radii on luminosity, means that we must have a very
precise knowledge of their temperature, and particularly their
luminosity (i.e., we need to know their parallaxes very well) to
determine their radii accurately. Fortunately, many of the red giants
on the Kepler target list will be bright enough to allow extraction of
individual oscillation frequencies for these stars, and not just an
average large spacing. As a result, it is unlikely that we will
have to use our method to determine red giant radii.

\item[--] Our tests show that the much-feared ``surface term'' should
not hamper the effort to determine stellar radii accurately from the
Kepler survey phase data. Tests with both error-less data and
simulations of what we can expect from Kepler show that the surface
term has a minimal effect on the inferred radius (at the level of
precision expected for the survey phase). While uncertainties in the
mixing length can cause errors, the effect is small, particularly for
main sequence stars.

\item[--] Finally, blind tests on three of the ``asteroFLAG cats'',
using data simulated to match the kinds of data expected from the
survey phase of Kepler, verify that it will be possible to infer
stellar radii successfully with our method to a typical precision of
just a few percent.

\end{itemize}

\acknowledgments

Work on the asteroFLAG simulator benefited from the support of the
International Space Science Institute (ISSI), through a workshop
programme award. This work was also partly supported by the European
Helio- and Asteroseismology Network (HELAS), a major international
collaboration funded by the European Commission's Sixth Framework
Programme. BiSON is funded by the UK Science and Technology Facilities
Council (STFC). WJC and YE acknowledge the support of STFC.

\clearpage

\clearpage


\end{document}